# Spin-dependent two-color Kapitza-Dirac effects

**S. McGregor, W. C.-W. Huang, B. A. Shadwick, and H. Batelaan**

Department of Physics and Astronomy, University of Nebraska—Lincoln, 208 Jorgensen Hall, Lincoln, Nebraska 68588-0299, USA
e-mail: hbatelaan2@unl.edu

**Abstract.** In this paper we present an analysis of the spin behavior of electrons propagating through a laser field. We present an experimentally realizable scenario in which spin dependent effects of the interaction between the laser and electrons are dominant. The laser interaction strength and incident electron velocity are in the non-relativistic domain. This analysis may thus lead to novel methods of creating and characterizing spin polarized non-relativistic femtosecond electron pulses.

## 1. Introduction

The capability to control electrons with laser light has been demonstrated with the higher light intensities that are provided by pulsed lasers[1, 2]. "In some of the first experiments, continuous electron beams were used. Consequently, most electrons were not affected by the light 1. More recently, pulsed electrons have also been affected by pulsed laser light[3, 4]. As more variations of pulsed electron sources that are synchronous with pulsed lasers are becoming available[5, 6], proposals have appeared that use such technology to control electron motion[7, 8]. As also table-top relativistic laser intensities are becoming more and more accessible, it is timely to consider the weaker interaction of electron spin with laser light.

Recently, it was predicted that X-ray laser light could be used to affect the electron spin of a beam of relativistic free electrons[9, 10]. More generally, electron spin control can provide an additional control to ultrafast electron diffraction[11, 12] and ultrafast electron microscopy[13, 14], similar to the non-pulsed version of spin-polarized low energy electron microscopy[15] (SPLEEM). For SPLEEM, GaAs polarized electron sources are used. However it is not clear what technology will be used for polarization control of femtosecond electron beams. In addition to its technological appeal, spin control may provide (through the spin-statistics connection) an opportunity to investigate quantum degeneracy in multi-electron pulses[16]. In view of these developments, we investigate the influence of visible light on the spin of non-relativistic electrons.

We report on an electron-laser configuration for which the spin dependent interaction is small, but dominant in the optical to near infrared domain. Specifically, a well collimated electron beam is intersected at right angles with two counter-propagating laser beams (Fig. 1) with frequencies $\omega$ and $2\omega$ ($\lambda = 2\pi c/\omega = 1\mu m$). Circular and linear polarizations of the two laser beams are considered. Circular polarization illustrates the spin coupling, while linear polarization, orthogonal to the electron beam propagation axis suppresses spin independent effects. For this configuration the regular Kapitza-Dirac effect[17] is absent due to the choice of widely separated frequencies, while the two-color Kapitza-Dirac effect[18] is absent because the electron velocity is chosen perpendicular to the laser polarization. The dominant interaction that remains is an interaction that scatters the electron beam by four momentum recoils and simultaneously flips the electron spin. We call this the spin-Kapitza-Dirac (SKD) effect. Ref.[9, 10] inspired the idea to use the orthogonality between the polarization and electron beam propagation axis. In this paper we use optical frequencies instead of X-rays, non-relativistic intensities instead of relativistic ones, and close to orthogonal angles between the electron beam and laser beams instead of angled ones. An extension, beyond the scope of the present paper, to relativistic intensities of the SKD in the optical frequency regime, or vice versa a study of the non-relativistic limit of Ref. [9,10] appears very interesting and may reveal other parameters ranges that are accessible to experiment.

The spin-flip probability for non-relativistic intensities is small, but detectable with current technology. The spin-flip probability increases for increasing intensity.

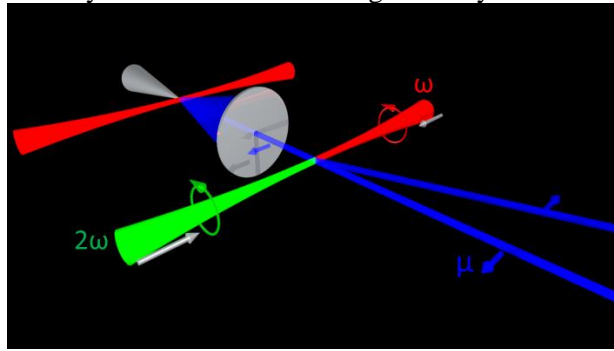

Figure 1. (color online) Schematic of the physical system. An electron pulse is generated from a field emission tip that is illuminated with a femtosecond laser[5]. The electron pulse is collimated (blue beam) and intersected with two counter propagating laser pulses of frequency $\omega$ (red) and $2\omega$ (green). Some electrons receive photon recoils of $4\hbar k$ while simultaneously flipping their spin (blue arrows) for appropriate chosen light polarization (see text for details).

A spin-dependent scattering could be used as an electron spin analyzer. No readily accessible techniques are available[19] to analyze the spin-polarization of a non-relativistic femtosecond electron pulse. Techniques for non-pulsed beams include Mott scattering[20], optical polarimetry[21], Rb spin-filter[22] and others. The most well-known and widely used Mott scattering requires currents exceeding 1 pA[23]. This current is usually not available for femtosecond electron pulses, so steady state methods do not easily transfer to pulsed scenarios. Relativistic and polarized pulses of electrons can be analyzed with Compton polarimetry[24]. However, the spin analyzing power drops off sharply with the relativistic gamma-factor. Femtosecond, nonrelativistic, pulsed, polarized electron sources are under development[25-27] and it is expected that analysis of their polarization will be needed. In general pulsed polarized electron sources are of interest for the broad area of spin physics[28].

The question may arise if such an optical control/analysis of electron spin is possible at all for non-relativistic electron motion. After all, Pauli pointed out that electrons cannot be polarized using ideas based on classical electron trajectories[29-35], as in a Stern-Gerlach device, even when the spin is treated quantum mechanically. This may appear to imply that the result obtained in this work could be ruled out based on a general principle. An earlier study based on classical mechanics for the same physical system as studied in this paper, indeed revealed no appreciable spin interaction,[36] consistent with Pauli's idea. However, our current analysis is fully quantum mechanical, thus circumventing the problem.

In section 2 non-relativistic perturbation theory is used to calculate the probability of the two-color spin-flip process. Additionally, processes that can potentially mask the process of interest, i.e. the regular two-color Kapitza-Dirac effect and a depolarizing effect, are calculated. It is shown that the spin dependent process can be dominant under the right conditions. In section 3 a systematic approach that discusses the order of magnitude of all possible perturbation terms is given. In section 4 a numerical solution to the Schrödinger equation is given that confirms the analysis of the previous sections. In section 5 a relativistic classical simulation is reported that confirms that for the studied scenarios all velocities stay well below c and the non-relativistic quantum approach is reasonable.

## 2. Perturbation Theory

The non-relativistic interaction Hamiltonian can be obtained by minimal substitution and considering the interaction of the electron dipole with the field[37],

$$H_{\text{int}} = -\frac{q}{2m}\left(\vec{p}\cdot\vec{A}+\vec{A}\cdot\vec{p}\right)+\frac{q^2 A^2}{2m}-\vec{\mu}\cdot\vec{B} \qquad (1)$$

Here, $q$ and $m$ are the electron charge and mass respectively, and the operators are given by, $\vec{p}$, the momentum, $\vec{A}$, the vector potential, $\vec{\mu}$, the electron magnetic moment, and $\vec{B}=\vec{\nabla}\times\vec{A}$, the magnetic field. This Hamiltonian can couple electron states with defined momentum and spin,

$$|\psi\rangle = |n\hbar k_z, \hbar k_x, m_S\rangle . \qquad (2)$$

The first entry in the above definition of the state vector of the electron defines the component of the electron momentum in the $z$-direction (i.e. aligned with the laser propagation direction), the second entry sets the initial electron momentum in the $x$-direction, and the third entry sets the magnetic quantum number corresponding to the projection of electron spin along the $z$-axis. The integer $n$ in the first entry is defined in anticipation that photon absorption and emission result in discrete changes of the electron momentum in terms of multiples of photon recoils, $\hbar k_z$. The Bragg condition leads to energy and momentum conservation for changes of the $z$-component of the electron momentum from $n\hbar k_z$ to $-n\hbar k_z$, while the x-component remains unchanged,[17] thus the quantum number $\hbar k_x$ is dropped from the state notation (Eq.2).

We investigate matrix elements that couple states as defined in Eq. 2 by the terms of the Hamiltonian Eq. 1. We will find that the $\frac{q^2 A^2}{2m}$ term in the Hamiltonian is responsible in time-dependent perturbation theory for the regular K-D effect[17], the terms $\frac{q^2 A^2}{2m}$ and $\frac{q}{m}\vec{p}\cdot\vec{A}$ together lead to the two-color K-D effect[18], consistent with the existing literature, while the terms $\frac{q^2 A^2}{2m}$ and $\vec{\mu}\cdot\vec{B}$ yield the spin dependent SKD effect that is the main focus of our current study.

Before starting the explicit calculation, it is useful to make some more general observations. Only processes which conserve kinetic energy in the laser field are considered in a perturbative approach. That this is valid is not obvious (neither is that the Bragg condition is always satisfied), and this needs to be justified. Below we report on a relativistic classical calculation that shows, for our parameters, that the change in the electron velocity along the direction of the laser propagation direction is limited to the order of a photon recoil. Our parameters are carefully chosen to avoid transverse acceleration and thus the weak spin-dependent scattering can become the dominant effect. Details of these choices are discussed below. The question whether or not an electron can be accelerated by laser fields has been debated for decades. In spite of the Lawson-Woodward theorem[38] it has been shown, that energy gain by laser interaction is possible for high energy electrons interacting with a tightly focused laser[39], and very recently even for approximately plane waves[40]. Our parameters do not satisfy the Lawson-Woodward criteria as the fields are not infinite in extent, the electron energy is not relativistic, and the ponderomotive potential is not negligible. The reason that the electron velocity, $v$, in the laser propagation direction changes little is that the electron and counter propagating laser pulses are timed such that the ponderomotive force from both pulses cancels. Our relativistic simulation does show that the longitudinal velocity can change significantly (see below).

To prevent a potentially dominant spin-independent scattering from overwhelming the weaker spin-dependent scattering, the physical parameters need to satisfy further criteria. At a laser intensity, $I=10^{19}$ W/m$^2$, and a wavelength, $\lambda= 800$ nm, an electron in a ponderomotive potential undergoes acceleration of up to $10^{22}$ m/s$^2$. The Larmor radiation rate at this acceleration, gives rise to a photon emission probability of $10^{-2}$ in an interaction time of $\tau=10$ ps. However, these photons are emitted in a large solid angle, give an average recoil in the laser propagation direction that is zero, and thus do not overwhelm the spin-dependent scattering.

We now continue with the explicit calculation of the spin-dependent perturbation term. In order to test whether or not spin-dependent scattering is plausible, perturbation theory was used to analyze each term in the interaction Hamiltonian in search of one term which would connect an initial spin state with a spin-flipped final state. For the purpose of this investigation we began with the vector potential corresponding with two circularly polarized laser beams which are counter-propagating along the $z$-axis,

$$\vec{A} = -\frac{A_0}{2\sqrt{2}} \exp\left(\frac{-t^2}{\tau^2}\right)\left(a_L e^{i(kz-\omega t)}(\hat{x}+i\hat{y}) + a_L^\dagger e^{-i(kz-\omega t)}(\hat{x}-i\hat{y})\right)$$
$$-\frac{A_0}{2\sqrt{2}} \exp\left(\frac{-t^2}{\tau^2}\right)\left(a_R e^{-i2(kz+\omega t)}(\hat{x}+i\hat{y}) + a_R^\dagger e^{i2(kz+\omega t)}(\hat{x}-i\hat{y})\right) \tag{3}$$

The choice of using raising and lowering photon number operators (with $[a,a^\dagger]=1$) is made to facilitate the selection of particular processes and is not essential. The calculations done in this section could have been done with classical fields to the same effect. The laser beam propagating in the direction of the positive $z$-axis has frequency $\omega$ and the laser propagating in the direction of the negative $z$-axis has frequency $2\omega$. Both laser beams have spin $\hbar$ in the direction of the positive $z$-axis. The magnetic dipole moment operator may be written in terms of the Pauli spin operator as $\vec{\mu} = \frac{-2\mu_B}{\hbar}\vec{S}$ where $\mu_B$ is the Bohr magneton. The $\frac{q}{m}\vec{p}\cdot\vec{A}$, $\frac{q^2 A^2}{2m}$, and $\vec{\mu}\cdot\vec{B}$ terms in the interaction Hamiltonian $H_{int}$ are

$$\frac{q}{2m}(\vec{p}\cdot\vec{A}+\vec{A}\cdot\vec{p}) = -\frac{qA_0}{2\sqrt{2}m}\exp\left(\frac{-t^2}{\tau^2}\right)\left(a_L e^{i(kz-\omega t)} + a_L^\dagger e^{-i(kz-\omega t)}\right)p_x$$
$$-\frac{qA_0}{2\sqrt{2}m}\exp\left(\frac{-t^2}{\tau^2}\right)\left(a_R e^{-i2(kz+\omega t)} + a_R^\dagger e^{i2(kz+\omega t)}\right)p_x, \tag{4}$$

$$\frac{q^2 A^2}{2m} = \frac{q^2 A_0^2}{8m}\exp\left(\frac{-2t^2}{\tau^2}\right)\left[a_L a_L^\dagger + a_R a_R^\dagger + a_L a_R^\dagger e^{i(3kz+\omega t)} + a_R a_L^\dagger e^{-i(3kz+\omega t)}\right]$$
$$+\frac{q^2 A_0^2}{8m}\exp\left(\frac{-2t^2}{\tau^2}\right)\left[a_L^\dagger a_L + a_R^\dagger a_R + a_L^\dagger a_R e^{-i(3kz+\omega t)} + a_R^\dagger a_L e^{i(3kz+\omega t)}\right], \tag{5}$$

$$\vec{\mu}\cdot\vec{B} = \frac{\mu_B A_0 k}{\sqrt{2}\hbar}\exp\left(\frac{-t^2}{\tau^2}\right)\left[\left(a_L e^{i(kz-\omega t)} - 2a_R e^{-i2(kz+\omega t)}\right)S_+ + \left(a_L^\dagger e^{-i(kz-\omega t)} - 2a_R^\dagger e^{i2(kz+\omega t)}\right)S_-\right], \tag{6}$$

where $S_+ = \frac{\hbar}{2}(\sigma_x + i\sigma_y)$ and $S_- = \frac{\hbar}{2}(\sigma_x - i\sigma_y)$ are the electron spin raising and lowering operators, respectively. The presence of the electron spin raising and lowering operators are a consequence of the choice of polarization. These operators can be used to connect initial and final states with different spin and therefore justify the choice of polarization in the search for spin-flip processes.

The first order probability amplitude is

$$C_{fi} = \frac{-i}{\hbar}\int_{-\infty}^{\infty} dt' H_{int}^{fi}(t'). \tag{7}$$

where $H_{int}^{fi} = \langle f|H_{int}|i\rangle$ couples the initial $|i\rangle$ to the final state $|f\rangle$. For spin-flip processes it is necessary to consider terms in the $\vec{\mu}\cdot\vec{B}$ part of the Hamiltonian. These contain the spin raising and lowering operators which are necessary to connect initial and final states with different spin in the matrix element. On examination of the $\vec{\mu} \cdot \vec{B}$ term it is apparent that such a first order process must be either single photon absorption or single photon emission because the terms in $\vec{\mu} \cdot \vec{B}$ each contain only one raising or lowering operator. Single photon processes are impossible because they cannot simultaneously conserve momentum and energy. It is therefore necessary to consider second order perturbation theory.

Using second order perturbation theory, the probability amplitude, $C_{fi}$, for transition between the initial ($i$) and final ($f$) states is found by summing over the intermediate state ($m$) for the 2$^{nd}$ and 3$^{rd}$ terms in the interaction Hamiltonian (Eq. 3,4)

$$C_{fi} = \frac{-1}{\hbar^2} \sum_m \int_{-\infty}^{\infty} dt' \int_{-\infty}^{t'} dt'' H_{int}^{fm}(t') H_{int}^{mi}(t''). \tag{8}$$

The matrix elements $H_{int}^{mi}$ and $H_{int}^{fm}$ correspond to transitions from the initial state to the intermediate state and from the intermediate state to the final state, respectively. For example, let us take $|N_\omega, N_{2\omega}, 2\hbar k, \uparrow\rangle$ and $|N_\omega + 2, N_{2\omega} - 1, -2\hbar k, \downarrow\rangle$ as initial and final states, respectively, where the photon state $|N_\omega, N_{2\omega}\rangle$ is labeled by the photon number for frequency $\omega$, $N_\omega$, and the photon number for frequency $2\omega$, $N_{2\omega}$. For the electron state $|2\hbar k, \uparrow\rangle$ the momentum label in the $x$-direction does not change and is dropped (cf. Eq. 2)). The electron wave function is a plane wave $\exp(i(\vec{k}_e \cdot \vec{x} - \omega_e t))$ where $\vec{k}_e$ and $\omega_e$ are the wave number and frequency of the electron, respectively. The $\frac{-\mu_B A_0 k}{\sqrt{2}\hbar} a_L^\dagger e^{-i(kz-\omega t)} S_-$ operator in the $\vec{\mu} \cdot \vec{B}$ term and the $\frac{q^2 A_0^2}{8m} a_R a_L^\dagger e^{-i(3kz+\omega t)}$ in the $\frac{q^2 A^2}{2m}$ term may be used to connect these two states. The matrix elements are given by

$$H_{int}^{mi}(t) = \frac{-\mu_B A_0 k}{\sqrt{2}\hbar} \exp\left(\frac{-t^2}{\tau^2}\right) \langle N_\omega + 1, N_{2\omega}, \hbar k, \downarrow | a_L^\dagger e^{-i(kz-\omega t)} S_- | N_\omega, N_{2\omega}, 2\hbar k, \uparrow\rangle$$
$$= \frac{-\mu_B A_0 k \sqrt{N+1}}{\sqrt{2}} \exp\left(\frac{-t^2}{\tau^2}\right) \exp(i(\omega_{mi} + \omega)t) \tag{9}$$

$$H_{int}^{fm}(t) = \frac{q^2 A_0^2}{8m} \exp\left(\frac{-2t^2}{\tau^2}\right) \langle N_\omega + 2, N_{2\omega} - 1, \hbar k, \downarrow | a_R a_L^\dagger e^{-i(3kz+\omega t)} | N_\omega + 1, N_{2\omega}, \hbar k, \downarrow\rangle$$
$$= \frac{q^2 A_0^2 \sqrt{N(N+2)}}{8m} \exp\left(\frac{-2t^2}{\tau^2}\right) \exp(i(\omega_{fm} - \omega)t), \tag{10}$$

where $N_{2\omega} = N_\omega = N$, $\omega_{mi} = \omega_m - \omega_i$ is the frequency difference between the initial and intermediate electron states, and $\omega_{fm} = \omega_f - \omega_m$ is the frequency difference between the intermediate and final electron states. The probability amplitude for this process may therefore be written as

$$C_{fi} = \frac{\mu_B q^2 k A_0^3 N^{3/2}}{8\sqrt{2} m \hbar^2} \int_{-\infty}^{\infty} dt' \int_{-\infty}^{t'} dt'' \exp\left[i\left((\omega_{fm} - \omega)t' + (\omega_{mi} + \omega)t''\right)\right], \tag{11}$$

where $\sqrt{N(N+1)(N+2)} \approx N^{3/2}$. It is apparent from this example that for the Hamiltonian given above there are only particular states that lead to a non-zero probability amplitude and identify the possible processes. Processes in which one of the laser pulses has no net change in photon number or processes in which the net change in photon number is identical for both pulses cannot simultaneously conserve momentum and energy[9]. Therefore, within the Bragg regime[17], spin flips are allowed for initial and final electron momentum states with $-2\hbar k$ and $2\hbar k$ using the $\vec{\mu} \cdot \vec{B}$ and $\frac{q^2 A^2}{2m}$ terms. All possible amplitudes corresponding to different intermediate states for processes involving a $4\hbar k$ momentum change with a spin flip from $\uparrow$ to $\downarrow$ are added together to determine the overall amplitude for the SKD process;

$$C_{fi} = \frac{\mu_B q^2 k A_0^3}{8\sqrt{2} m\hbar^3} \int_{-\infty}^{\infty} dt' \int_{-\infty}^{t'} dt'' \langle N_\omega + 2, N_{2\omega} - 1, -2\hbar k, \downarrow | a_R a_L^\dagger e^{-i(3kz+\omega t)} | N_\omega + 1, N_{2\omega}, \hbar k, \downarrow \rangle \exp(i\omega_{fm} t') \exp\left(\frac{-2t'^2}{\tau^2}\right)$$

$$\times \langle N_\omega + 1, N_{2\omega}, \hbar k, \downarrow | a_L^\dagger e^{-i(kz-\omega t)} S_- | N_\omega, N_{2\omega}, 2\hbar k, \uparrow \rangle \exp(i\omega_{mi} t'') \exp\left(\frac{-t''^2}{\tau^2}\right)$$

$$+ \frac{\mu_B q^2 k A_0^3}{8\sqrt{2} m\hbar^3} \int_{-\infty}^{\infty} dt' \int_{-\infty}^{t'} dt'' \langle N_\omega + 2, N_{2\omega} - 1, -2\hbar k, \downarrow | a_L^\dagger a_R e^{-i(3kz+\omega t)} | N_\omega + 1, N_{2\omega}, \hbar k, \downarrow \rangle \exp(i\omega_{fm} t') \exp\left(\frac{-2t'^2}{\tau^2}\right)$$

$$\times \langle N_\omega + 1, N_{2\omega}, \hbar k, \downarrow | a_L^\dagger e^{-i(kz-\omega t)} S_- | N_\omega, N_{2\omega}, 2\hbar k, \uparrow \rangle \exp(i\omega_{mi} t'') \exp\left(\frac{-t''^2}{\tau^2}\right)$$

$$+ \frac{\mu_B q^2 k A_0^3}{8\sqrt{2} m\hbar^3} \int_{-\infty}^{\infty} dt' \int_{-\infty}^{t'} dt'' \langle N_\omega + 2, N_{2\omega} - 1, -2\hbar k, \downarrow | a_L^\dagger e^{-i(kz-\omega t)} S_- | N_\omega + 1, N_{2\omega} - 1, -\hbar k, \uparrow \rangle \exp(i\omega_{fm} t') \exp\left(\frac{-t'^2}{\tau^2}\right)$$

$$\times \langle N_\omega + 1, N_{2\omega} - 1, -\hbar k, \uparrow | a_R a_L^\dagger e^{-i(3kz+\omega t)} | N_\omega, N_{2\omega}, 2\hbar k, \uparrow \rangle \exp(i\omega_{mi} t'') \exp\left(\frac{-2t''^2}{\tau^2}\right)$$

$$+ \frac{\mu_B q^2 k A_0^3}{8\sqrt{2} m\hbar^3} \int_{-\infty}^{\infty} dt' \int_{-\infty}^{t'} dt'' \langle N_\omega + 2, N_{2\omega} - 1, -2\hbar k, \downarrow | a_L^\dagger e^{-i(kz-\omega t)} S_- | N_\omega + 1, N_{2\omega} - 1, -\hbar k, \uparrow \rangle \exp(i\omega_{fm} t') \exp\left(\frac{-t'^2}{\tau^2}\right)$$

$$\times \langle N_\omega + 1, N_{2\omega} - 1, -\hbar k, \uparrow | a_L^\dagger a_R e^{-i(3kz+\omega t)} | N_\omega, N_{2\omega}, 2\hbar k, \uparrow \rangle \exp(i\omega_{mi} t'') \exp\left(\frac{-2t''^2}{\tau^2}\right) \quad (12)$$

The integrals were calculated numerically using Maple and the results are shown in column 2 of Table 1. Also see section 3 for an analytic comparison. Throughout a laser focus width of $d=100$ $\mu m$ was used. This leads to an approximate photon number given by $N = I\pi d^2 / 4\hbar\omega$. Similar to the integral in Eq. 12, there are two integrals representative of processes by which the electron can receive a spin flip from $\uparrow$ to $\downarrow$ with no net momentum kick from only one of the lasers that must be summed coherently. Such an event may flip a spin of an electron that already received a momentum kick and spin flip, and thus undo the effect we are interested in. The results of the calculations are shown in column 1 of Table 1.

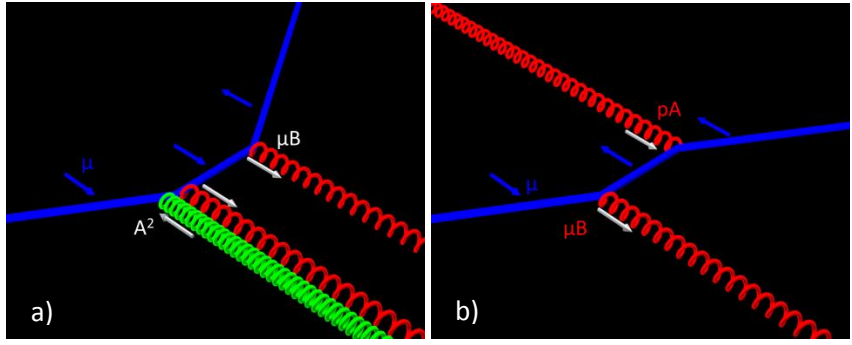

Figure 2. (color online) Spin dependent processes. a) An example is shown of three photon process by which the electron receives a spin flip and a momentum kick by absorbing one 2ω photon and emitting two ω photons. At the first vertex (form left to right) an absorption and emission of a 2ω photon and a 1ω photon is shown, respectively, indicating the use of the $A^2$ term of the Hamiltonian and no spin-flip. The second vertex shows an emission of a 1ω photon, indicating the use of the μB term of the Hamiltonian, accompanied with a spin-flip. b) An example is shown of two photon process by which the electron receives a spin flip without an overall deflection by emitting and absorbing photons from the same laser. At the first vertex an emission of a photon is shown, in this case associated with the use of the μB term of the Hamiltonian. The second process shown represents the absorption of a photon, by, for example, the use of the pA term of the Hamiltonian.

|  | Depolarizer | SKD effect | Two color K-D effect |
|---|---|---|---|
| Intensity $I$ | $10^{18} W/m^2$ | $10^{18} W/m^2$ | $10^{15} W/m^2$ |
| Velocity $v$ | $10^7 m/s$ | $10^7 m/s$ | $10^7 m/s$ |
| Wavelength $\lambda$ | $1064 nm$ | $1064 nm$ | $1064 nm$ |
| Interaction Time $\tau$ | $100 ps$ | $100 ps$ | $100 ps$ |
| Probability $P$ | 0.00576 | 0.00128 | $7.4 \times 10^{-4}$ |
| Scaling | $P = \alpha I^2 v^2 \lambda^2 \tau^2$ | $P = \alpha I^3 \lambda^4 \tau^2$ | $P = \alpha I^3 v^2 \lambda^6 \tau^2$ $P = \alpha I^3 v^2 \lambda^6 \tau^2$ |
| $\alpha$ | $5.09 \times 10^{-21}$ | $9.96 \times 10^{-14}$ | $5.12 \times 10^{-7}$ |

Table 1. Probabilities. The probabilities of a spin flip which acts as a depolarizer (column 1), a spin flip which is accompanied with a momentum kick, i.e. the SKD effect (column 2), and the two color K-D effect (column 3) are given for an example of laser intensity, electron velocity, laser wavelength, and interaction time.

Given the numbers in Table 1, it appears that an interaction in which an electron spin flip due to laser interaction is possible but these are only representative of a relatively small number of potentially relevant scattering events that may take place in the physical scenario described above. With only this information we cannot know that the spin dependent effect is dominant. It is therefore necessary to compute the spin flip probability in a manner which incorporates all possible interactions described by the Hamiltonian and conceive of a physical scenario in which a spin flip is dominant.

### 3. Relative orders of magnitude

In the previous section the focus was on specially selected perturbative terms that are important to our discussion. Here a more systematic approach is followed that includes relative order of magnitude estimations. Ignoring specific choices of the physical parameters, in first order perturbation theory three matrix elements $H_{fi}^j \equiv \langle f | H^j | i \rangle$ are possible (see Eq. 5), where the operators are $H^1 = q^2 A^2 / 2m$, $H^2 = q \vec{p} \cdot \vec{A} / m$, and $H^3 = \vec{\mu} \cdot \vec{B}$. At this point we consider, as before, two counter propagating laser pulses that are intersected with an electron pulse, and the frequency of the fields is given by $\omega_1$ and $\omega_2$. The probability amplitude (Eq.6) is rewritten as $C_{fi}^{j} \equiv \|H^{j}\| f_{fi}^{j}(\tau)$, where the magnitude in decreasing order will turn out to be given by $\|H^1\| \equiv q^2 A_0^2 / 2m$, $\|H^2\| \equiv q\vec{p} \cdot \vec{A}_0 / m$, and $\|H^3\| \equiv \vec{\mu} \cdot \vec{B}_0 / m$, with $B_0 = kA_0$ $B_0 = kA_0$. The value of the amplitude (Eq. 5) can be approximated (see Appendix) by

$$C_{fi}^{j} \approx \|H^{j}\| \tau / \hbar. \tag{13}$$

The amplitude $C_{fi}^{j=1}$ is non-zero for $\omega_1 = \omega_2$ with an initial and final state choice of $-\hbar k$ and $\hbar k$. This process is the well-known KD-effect[17], conserves energy and momentum, and is a two-photon process. The number of photons in a process can be recognized by inspecting the power of the field. From equation (11) the probability of scattering is given by $(q^2 A_0^2 \tau / 2m\hbar)^2$ in agreement with previous work[2,17].

Energy and momentum can also be conserved for $C_{fi}^{j=1}$ when $\omega_1 \neq \omega_2$. However, when $\omega_1 = 2\omega_2$, for example, the electron needs to move relativistically at steep angles with respect to the laser propagation direction. The amplitudes $C^{j=2}$ and $C^{j=3}$ involve the interaction with one photon, which is kinematically not allowed.

In second order perturbation theory, combinations of two terms of $H^j$ need to be considered. The matrix elements $H_{fmi}^{jj'} \equiv \langle f | H^j | m \rangle \langle m | H^{j'} | i \rangle$ give rise to a probability amplitude $C_{fmi}^{jj'} \equiv \|H^j\| \|H^{j'}\| g_{fmi}^{jj'}(\tau)$. The value of the amplitude (using Eq. 9) can be approximated (see Appendix) by

$$C_{fmi}^{jj'} \equiv \|H^j\| \|H^{j'}\| \frac{\tau}{\omega \hbar^2} \frac{\hbar k}{mc}. \tag{14}$$

The term $C_{fmi}^{j=1,j'=1}$ for $\omega_1 = 2\omega_2$ (where $\omega_1$ comes from one direction and $\omega_2$ from the other (see figure 1)) does not conserve energy and momentum, unless the initial and final electron state are identical. It is thus possible that our wanted spin-dependent kick is followed by this process. However, this term does not couple spin or momentum and will not dilute our process of interest.

The second order term $C_{fmi}^{j=1,j'=2}$ for $\omega_1 = 2\omega_2$ is the regular two-color KD-effect[18]. From equation (12) the probability of scattering is given by $(kq^3 A_0^2 \vec{p} \cdot \vec{A}_0 \tau / 2m^3 \omega \hbar c)^2$ in agreement with previous work[18]. To suppress this term, $\vec{p}$ is chosen perpendicular to $\vec{A}_0$ (The required accuracy of the angle is discussed in section 6). This also implies that $C_{fmi}^{j,j'=2} = 0$. The next term to consider is $C_{fmi}^{j=1,j'=3}$. That is the term of interest of this paper (see the derivation in the previous section). The last second order perturbative term, $C_{fmi}^{j=3,j'=3}$, can only conserve energy and momentum when the momentum and spin state is unchanged, and thus will not be observable in a scattering experiment.

Higher order processes are worth considering as well, despite the fact that it seems likely that they will be negligible compared to the spin dependent process of interest. For example, third order perturbation theory might be expected to result weaker processes than lower order perturbative processes, however, the combination of three strong matrix elements (i.e. matrix elements computed from the $q^2 A^2 / 2m$ term of the Hamiltonian) might provide stronger scattering than the spin-dependent scattering term of interest here, that has one strong and one weak matrix element. To consider the effects of all higher order processes a numerical integration of the Schrödinger equation was performed.

## 4. Numerical integration of the Schrödinger equation.

The purpose of the numerical simulation is to verify that the perturbative approach is sufficient, and, for example, the inclusion of third order perturbative terms is indeed not required. We calculate the electron scattering to different states of momentum and spin by numerically solving the Pauli equation. The electron state is a plane wave described by

$$|\psi(t=0)\rangle = |\hbar k_{z0}, \hbar k_x, m_s\rangle, \quad (15)$$

where $\hbar k_{z0}$ is the initial transverse momentum. The electron then passes through the two-color light $\vec{A}(\vec{z},t) = \vec{A}_R(\vec{z},t) + \vec{A}_L(\vec{z},t)$, which is composed of two light fields coming from opposite directions,

$$\vec{A}_L(\vec{z},t) = 2A_L e^{-(t/z)^2} \cos(k_L z - \omega_L t)\hat{\varepsilon}_L$$
$$\vec{A}_R(\vec{z},t) = 2A_R e^{-(t/z)^2} \cos(k_L z + \omega_L t)\hat{\varepsilon}_L \quad (16)$$

The frequencies chosen are again $\omega_L = \omega$ and $\omega_R = 2\omega$. The field polarization is described by the unit vector $\hat{\varepsilon}$ in the $x-y$ plane. In the above perturbative calculation, the light was chosen to be circularly polarized. Here we chose linearly polarized light, because the contribution from the $q\vec{p} \cdot \vec{A}/m$ term needs to be controlled. As in the perturbative calculation, the light field has no spatial dependence in the $x$-direction, and the electron momentum changes in photon recoil increments, $\hbar\omega/c$, while the $|k_x\rangle$ state stays unchanged,

$$|\psi(t)\rangle = \sum_{n,j} C_{n,j}(t) e^{i\omega_n t} |\hbar k_{z0} + \hbar k_n, \hbar k_x, m_{s,j}\rangle, \quad (17)$$

where $\omega_n = \hbar\left[k_x^2 + (k_{z0} + k_n)^2\right]/2m$, and $k_n = nk = n\omega/c$. The electron state (Eq. 17) has been generalized as compared to Eq. 2 in the sense that the initial momentum $\hbar k_{z0}$ is not limited to multiples of photon recoil. In order to calculate the amplitudes $C_{n,s}$, we solve for the Pauli equation, using the Hamiltonian,

$$H = H_0 + H', \quad (18)$$

which has been decomposed into an unperturbed part,

$$H_0 = \frac{p_x^2}{2m} + \frac{p_y^2}{2m}, \quad (19)$$

and a perturbed part,

$$H' = -\frac{q}{m}A_x p_x + \frac{q^2}{2m}\left(A_x^2 + A_y^2\right) - \frac{\hbar q}{2m}\left(B_x \sigma_x + B_y \sigma_y\right), \tag{20}$$

Where $\sigma_i$ are the Pauli matrices. The Pauli equation can now be written as

$$\frac{d}{dt}C_{m,i}(t) = -\frac{i}{\hbar}\sum_{n,j} H'_{2(m-1)+i,2(n-1)+j} C_{n,j}(t) e^{i\omega_{mn}t} \tag{21}$$

where $\omega_{mn} = \omega_m - \omega_n$ and

$$H'_{2(m-1)+i,2(n-1)+j} = \langle \hbar k_m, \hbar k_x, m_{s,i} | H' | \hbar k_n, \hbar k_x, m_{s,j}\rangle$$
$$= -\frac{q\hbar k_x}{m}\langle \hbar k_m, \hbar k_x | A_x | \hbar k_n, \hbar k_x\rangle\langle m_{s,i} | m_{s,j}\rangle + \frac{q^2}{2m}\langle \hbar k_m, \hbar k_x |(A_x^2 + A_y^2)| \hbar k_n, \hbar k_x\rangle\langle m_{s,i} | m_{s,j}\rangle$$
$$-\frac{\hbar q}{2m}\left(\langle \hbar k_m, \hbar k_x | B_x | \hbar k_n, \hbar k_x\rangle\langle m_{s,i} | \sigma_x | m_{s,j}\rangle + \langle \hbar k_m, \hbar k_x | B_y | \hbar k_n, \hbar k_x\rangle\langle m_{s,i} | \sigma_y | m_{s,j}\rangle\right). \tag{22}$$

An example of a numerical result is given in Figure 3. The initial electron state was chosen to be $\hbar k_z = 2\hbar k$ and $|m_s\rangle = |\uparrow\rangle$. The initial electron velocity was $10^7$ m/s and the laser pulses were polarized in the $y$-direction. The probability of the spin SKD processes are compared with the perturbation calculation (Figure 3) showing agreement between the two methods. The probability of the spin dependent depolarizing effect (Figure 2b) is also shown in Figure 3. The two color KD effect and as well as the regular KD effect (for $\omega_L = \omega_R$) are shown for comparison.

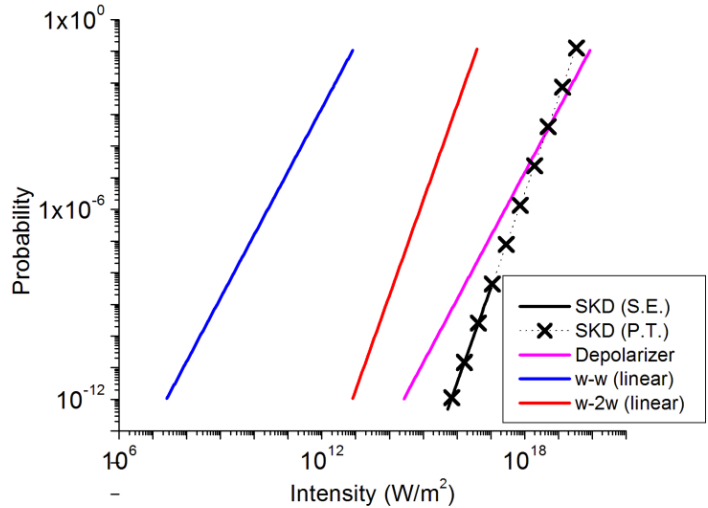

Figure 3: (color online) Intensity dependence. The probability of the SKD process as computed by numerical integration of the Schrodinger equation (S.E.), the SKD probability as obtained with perturbation theory (P.T.) are shown in figure 3 to be in agreement. Additionally, the two-color KD (ω-2ω) effect, the regular KD (ω-ω) effect (for $\omega_L = \omega_R$), and the depolarizer are shown for comparison. The SKD and two-color KD have a slope of three, indicating a three photon process, while the depolarizer and the regular KD process have a slope of two.

The probability of the SKD process is about 0.01 at $10^{19}$ W/m$^2$. The depolarizing process is weaker by about an order of magnitude (i.e. of the electrons which undergo the SKD process, only approximately 1 in $10^3$ will return to spin up).

The probability associated with final momentum states having $p_z = n\hbar k_0$ for $n = -7$ through 7 are shown in Figure 4 for spin up and spin down. These values were computed for a laser intensity of $10^{18}$ W/m$^2$.

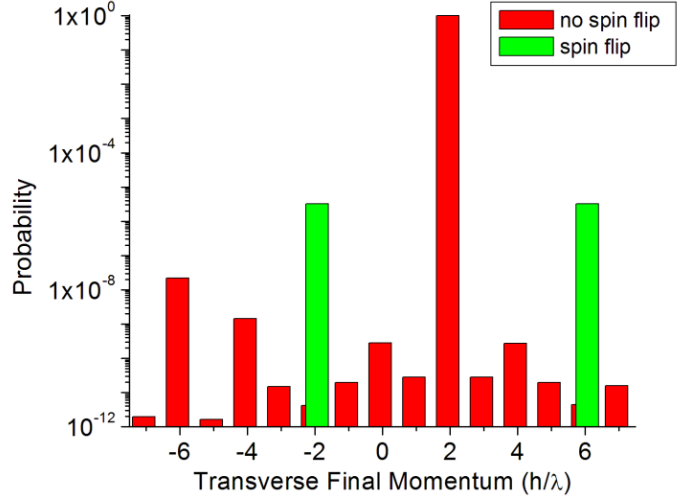

Figure 4: (color online) Momentum distribution of the SKD process. The initial state is spin up with $p_z = 2\hbar k$. The probability of the final electron momentum states having $z$-components of $p_z = n\hbar k = nh/\lambda$ for $n = -7$ through 7 are shown for spin up (dark red) and spin down (light green).

At this modest intensity the initial state at $2\hbar k$ is mostly unaffected. The largest probability diffraction occurs into the states with -2 and $6\hbar k$. The SKD process that satisfies the Bragg condition is the one with the same electron kinetic energy (-2 $\hbar k$). At $10^{18}$W/m$^2$ the probability of the spin flip kick is about $10^{-6}$ (cf. figure 3). Because the interaction time is chosen to be in the diffractive regime (for which the electron kinetic energy in a one-dimensional calculation is not conserved[17]), symmetric diffraction to 6 $\hbar k$ occurs. As usual asymmetric diffraction occurs in the Bragg regime. The peak at -6 $\hbar k$ is not a sequence of SKD processes as the probability would have to be about $(10^{-6})^2$.

It should be note that the field (Eq. 16) is a plane wave with infinite extension in all spatial directions. Thus the effect of spatial gradients has not been addressed up to now. Additionally, both the perturbative and numerical approaches are non-relativistic, while intensities of $10^{19}$ W/m$^2$ might lead to relativistic velocities of the electron in the laser light. To assess the effect of spatial gradients and the validity of a non-relativistic approach a relativistic classical simulation is performed.

## 5. Relativistic classical simulation.

Classical effects that can invalidate the above are: *i*) the electrons reflect from the ponderomotive barrier presented to the electron by the laser light, *ii*) the electrons reach relativistic factors $\gamma$ that strongly exceed 1, and *iii*) the electrons are deflected transversely by much more than the deflection produced by the spin-dependent scattering (i.e., four photon recoils).

Predictions that have been made in the previous sections were based on non-relativistic quantum mechanics. This requires sufficiently low velocity electrons throughout the interaction with the laser field. Additionally scattering from the ponderomotive potential will result in broadening of the diffraction peaks. If the maximum deflection due to ponderomotive scattering exceeds that of the spin dependent scattering, the peak corresponding to the effect of interest will be resolved. Finally, if the electron is reflected back from whence it came, it cannot pass through the laser and arrive at the detector.

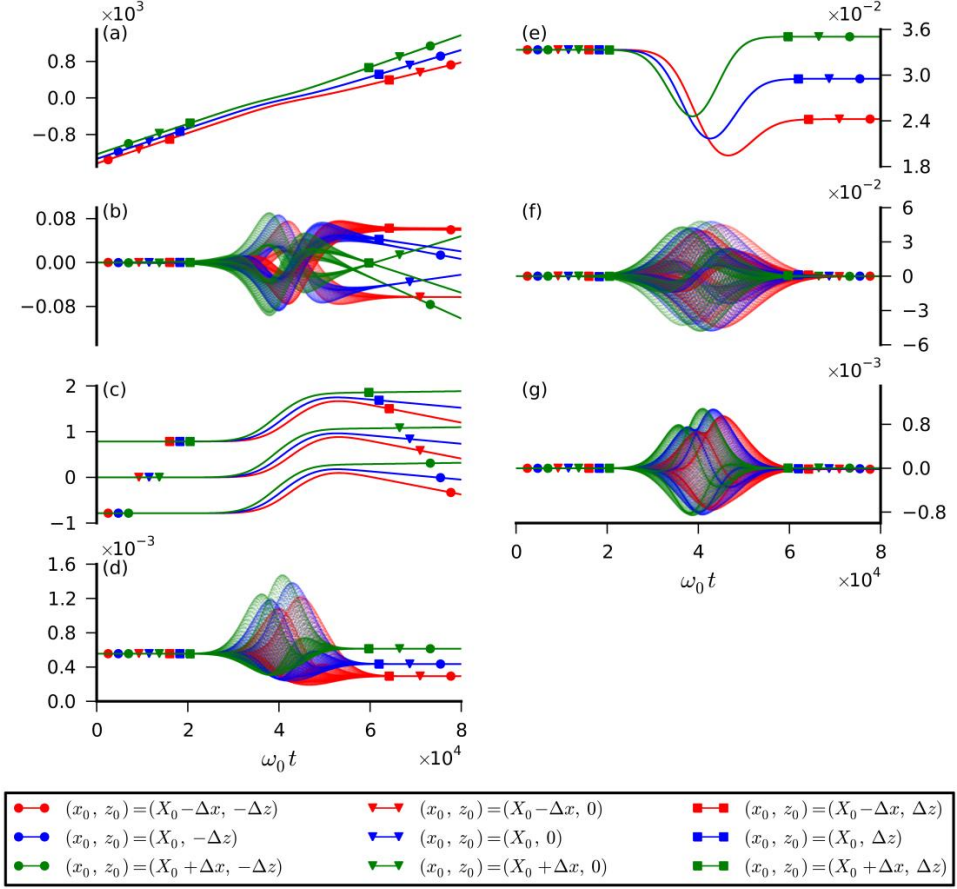

Figure 5: (color online) Shown here is the time dependence of the position of the electron a) $k_0 x$, b) $k_0 y$, c) $k_0 z$; momentum of the electron e) $p_x / mc$, f) $p_y / mc$, g) $p_z / mc$; and d) the relativistic factor $\gamma - 1$. Each is shown for different initial positions. See text for details.

The relativistic equations of motion are solved numerically for a single electron traversing counter-propagating laser pulses. The electron momentum and position evolve according to

$$\frac{d\vec{p}}{dt} = q\left(\vec{E} + \vec{v} \times \vec{B}\right)$$
$$\frac{d\vec{r}}{dt} = \vec{v},$$
(23)

where $\vec{p} = \gamma m \vec{v}$, $\gamma^2 = 1 + p^2/m^2c^2$, and the electric and magnetic fields are evaluated at the location of the electron. The laser pulses, taken to be described by the lowest order paraxial Gaussian mode[41], are polarized in the $y$-direction, propagate in the $z$-direction and have a 100 μm spot size at the focus. The pulse propagating in the positive $z$-direction has frequency $\omega_0$ corresponding to a wavelength of 1μm with a peak value of the vector potential given by $qA/mc = 0.03$ $I_{\omega_0} = 1.24 \times 10^{19} W/m^2$ while the pulse propagating in the negative $z$-direction has frequency $2\omega_0$ with peak value of the vector potential given by $qA/mc = 0.02$ $I_{2\omega_0} = 2.20 \times 10^{19} W/m^2$. For both laser pulses, the vector potential has the Gaussian temporal profile $\exp\left[-(z-ct)^2/\tau^2\right]$ with $\tau = 10 ps$. The laser pulses are initialized such that they reach the focus at $z=0$ at $\omega_0 t = 4000$. The electron is initially propagating in the positive $x$-direction with a velocity $v_0 = c/30$. The sensitivity of the deflection to initial conditions can be seen by examining trajectories over a set of initial conditions (see Figure 1). Initially, we take $y=0$ and $(x,z)$ from the set of nine pairs $[X_0 - \Delta x, X_0, X_0 + \Delta x] \times [-\Delta z, 0, \Delta z]$, where $k_0 X_0 = 4000 v_0/c$, $k_0 \Delta x = 100$, and $k_0 \Delta z = \pi/4$. The value for $X_0$ is chosen such that, in the absence of an interaction with the laser field, the electron would arrive at the origin at the same instant that the laser pulses reach focus and have maximal overlap. The value of $\Delta x$ is chosen to be comparable the laser spot size, and $\Delta z$ is chosen comparable to the laser wavelength. All computations are performed in dimensionless variables using $\omega_0$ and $k_0 = \omega_0/c$ to set the temporal and spatial scales while $mc$ is used for the momentum scale.

The top three panels in the left (right) column of Figure 5 indicate the electron position (momentum) as is propagates through the laser pulses. Panel (b) and (f) show that when the electron is present in the laser field it performs an oscillatory motion, which is due to the laser electric field. Panel (a) and (e) show that the ponderomotive potential affects the electron motion in the forward direction, but does not reflect from the barrier (assumption *i*)). Panel (g) shows that the magnetic part of the Lorentz force causes an oscillatory motion. Panel (d) shows that the gamma factor does not strongly deviate from one at any time (assumption *ii*). The variation in the slope of the outgoing electron in panel (c) shows that the transverse velocity reaches maximum values close to $4\hbar k/m$ (assumption *iii*).

From this analysis it is possible to deduce what the limitations are in a demonstration of the spin dependent effect. While the intensity of the lasers is not limited by the demand of keeping the electron trajectory non-relativistic it is limited by deflection. While the transverse ponderomotive scattering in this case is sufficiently low an increase in intensity would lead to increased deflection pushing the broadening of diffraction peaks to an unacceptable level.

## 6. Discussion

It is perhaps curious that a classical calculation using the Bargmann-Michel-Telegdi equations leads to vanishingly small spin-flip probabilities for the same physical configuration and parameters as used in the present analysis[36]. This is especially so, given that the regular Kapitza-Dirac and two-color Kapitza-Dirac effect can both be analyzed classically and quantum-mechanically to give a similar size effect[17]. We speculate that the current effect is a true quantum effect as it apparently has a zero classical counter-part. Pauli provided a proof that the design of a device that completely analyzes the spin of an electron, such as an electron Stern-Gerlach device is not possible based on the concept of classical trajectories[29, 30]. However, this principle can be side-stepped by a design motivated by quantum mechanical principles. This has been shown for the electron Stern-Gerlach magnet[35]. The current effect appears to fall into the same category. An incoming unpolarized electron beam could be analyzed completely according to its spin state.

It appears there is a window of parameter values where spin-dependent scattering of laser light with electrons is dominant. However, in a real experiment spurious effects can be present and overwhelm the process of interest. Three of such effects are now discussed. With short pulses the frequency distribution of one laser beam (centered around $\omega$) could be broadened so that it has a nonzero value at the peak of the distribution of the counter-propagating laser beam (centered around $2\omega$). Since the regular ($\omega-\omega$) K-D effect[2] is so much stronger than the effects considered in this paper, one may wonder if it will overshadow our effect in spite of the fact that the two frequencies are an octave apart. If 10 ps pulses of light with 1064nm wavelength are used, than the difference between the two frequencies is about $10^4$ times the width of each distribution. This leads to negligible effect for a Lorentzian (or Gaussian) spectral distribution of the laser. The regular K-D effect is thus sufficiently reduced by the separation of the frequencies.

In practice, the $2\omega$ laser beam may be generated by up-conversion and result two co-propagating beams that need to be separated optically. If this is not done the regular K-D effect will still be present. Dichroic mirrors and filtering can be used to provide separation of the two frequencies. Our analysis indicates that the ratio of the first order over a second order process (Eq. 11 and 12) is given by $Z = z_\mu - z_q$. For the spin dependent coupling $\|H^{j'}\| \approx \mu B$ and an intensity of $10^{19}$ W/m$^2$ this is about $10^6$. To suppress the regular K-D effect by this much an isolation in intensity of $10^{-6}$ is thus required.

The strong regular two-color K-D effect is suppressed by the choice that the laser polarization is perpendicular to the electron velocity, because this K-D effect has $\|H^{j'}\| \approx \frac{q}{m}\vec{p}\cdot\vec{A}$ term in the Hamiltonian. However the polarization angle or electron beam direction may be misaligned. The ratio of the regular two-color K-D effect over the spin-dependent K-D effect is $\frac{q}{m}\vec{p}\cdot\vec{A}/\mu B$, which equals about $10^5$. Since the amplitude of the regular effect is proportional to $\cos(\theta)$, where $\theta$ is the angle between the electron velocity and the laser polarization, than angle should be aligned better than $0.01$ mrad from the perpendicular.

The three spurious effects given above can be discriminated against as they have distinguishing features which can isolate them from the spin-dependent scattering term of interest. The spin-dependent effect is not velocity dependent nor polarization angle dependent in contrast to the two color K-D effect. It can also be distinguished from the regular K-D effect by the different intensity dependence.

It is important to note that the effect discussed in this paper differs from the relativistic effect proposed by Ahrens et al.[9]. In the paper by Ahrens et al.[9] the frequency of the two laser beams is the same, the laser light has a photon energy of 3.1 keV, and the 176 keV electrons are incident at an angle that is far from perpendicular to the lasers.

Given the wavelength dependence of the two and three photon effects it is tempting to consider lowering the frequency of the lasers to dramatically boost the probability. If the wavelength is increased the focal width too will increase which eventually will result in a wavelength dependent interaction time. Assuming an interaction time that is proportional to wavelength, the two and three photon effects become proportional to $\lambda^4$ and $\lambda^6$ respectively. While the ratio of the probabilities remains the same in this case the two effects become more strongly wavelength dependent by an added factor of $\lambda^2$ thus increasing the benefit of a longer wavelength.

It is apparent from the numbers presented in Table 1 that with the right parameters the probabilities of the two photon and three photon effects are comparable. Since the probability of a spin-flip with no momentum kick due to the two photon process is the same for both spin states regardless of input angle this effect can be thought of as a depolarizer. If a polarized beam of electrons propagates through a laser field some of the electrons will not flip, some will flip once, while others will flip more than once. The output electron beam will be depolarized to some extent which depends on the intensity of the laser field. This could potentially be a problem. If the three photon process is used to create a polarized electron beam, that beam could be depolarized by the very same set of counter propagating lasers before it has a chance to exit the field. With such an experiment in mind, it is therefore necessary to set the parameters such that the probability associated with the two photon process is small compared to the probability associated with the three photon process.

## 7. Conclusion

In this paper we have shown that a dominant spin dependent K-D effect is possible, given the appropriate laser configuration. As compared to the interesting recent work of Ahrens et al[9, 10], the current work extends spin-control of electron by light into the non-relativistic and visible light domain. This effect could be used as an ultrafast spin polarized electron source or to analyze such a source. Applications include polarization control for ultrafast electron diffraction and ultrafast electron microscopy, as well as more fundamental physics studies such as the effect of the Pauli exclusion principle on the propagation of multi-electron pulses[16].

**Acknowledgements**
This work was supported by the National Science Foundation under grant no. 0969506 and 1306565.

**Appendix**

In order to calculate the approximations given in section direct integration of the probability amplitude was performed. In the case of the regular K-D effect calculation of the integral shown in equation (5) was performed. The matrix element chosen corresponds to the $\frac{q^2 A^2}{2m}$ term in the Hamiltonian where $\vec{A}$ is the vector potential corresponding to two counter propagating lasers of frequency $\omega$ given by

$$\vec{A} = \frac{A_0}{2} \exp\left(\frac{-t^2}{\tau^2}\right)\left(a_L e^{i(kz-\omega t)} + a_L^\dagger e^{-i(kz-\omega t)} + a_R e^{-i(kz-\omega t)} + a_R^\dagger e^{i(kz+\omega t)}\right)\hat{x} \quad . \tag{A1}$$

Taking the operator $\frac{q^2 A_0^2}{2m} a_L^\dagger a_R \exp\left(\frac{-2t^2}{\tau^2} - i2kz\right)$ in the $\frac{q^2 A^2}{2m}$ term in the Hamiltonian which is descriptive of a $-2\hbar k$ momentum kick and applying equation (5) gives

$$C_{fi} = \frac{-i}{\hbar} \int_{-\infty}^{\infty} H_{\text{int}}^{fi}(t')dt' = \frac{-iq^2 A_0^2}{2m\hbar} \int_{-\infty}^{\infty} \langle N+1, N-1, -\hbar k | a_L^\dagger a_R \exp(-i2kz) | N, N, \hbar k \rangle \exp\left(\frac{-2t'^2}{\tau^2}\right)dt'$$

$$= \frac{-iq^2 A_0^2 N}{2m\hbar} \int_{-\infty}^{\infty} \exp\left(\frac{-2t'^2}{\tau^2} + i\omega_{fi} t'\right)dt' \tag{A2}$$

Since the initial and final state of the electron satisfy the Bragg condition, the frequency difference between the two is zero $(\omega_{fi} = 0)$.

$$C_{fi} = \frac{-iq^2 A_0^2 N}{2m\hbar} \int_{-\infty}^{\infty} \exp\left(\frac{-2t'^2}{\tau^2}\right) dt' = -\sqrt{\frac{\pi}{2}} \frac{iq^2 A_0^2 N\tau}{2m\hbar} = -\sqrt{\frac{\pi}{2}} \frac{iq^2 I\tau}{\hbar mc\varepsilon_0 \omega^2}, \quad (A3)$$

where $I$ is the laser intensity. For the two-color K-D effect the integral shown in equation (6) was performed. The matrix elements chosen corresponds to the $\frac{q^2 A^2}{2m}$ and $\frac{q}{m}\vec{p}\cdot\vec{A}$ terms in the Hamiltonian where $\vec{A}$ is the vector potential corresponding to two counter propagating lasers of frequencies $\omega$ and $2\omega$.

$$\vec{A} = \frac{A_0}{2} \exp\left(\frac{-t^2}{\tau^2}\right)\left(a_L e^{i(kz-\omega t)} + a_L^\dagger e^{-i(kz-\omega t)} + a_R e^{-i2(kz-\omega t)} + a_R^\dagger e^{i2(kz+\omega t)}\right)\hat{x}. \quad (A4)$$

Accounting for all possible combinations of operators contained in the $\frac{q^2 A^2}{2m}$ and $\frac{q}{m}\vec{p}\cdot\vec{A}$ terms which give rise to a momentum kick of $-4\hbar k$ results in the probability amplitude

$$C_{fi} = \frac{-q^3 A_0^3 N^{\frac{3}{2}} p_x}{16m^2\hbar^2}\left\{\int_{-\infty}^{\infty} e^{i\left[\left(\frac{2\hbar k^2}{m}+2\omega\right)t'\right]} e^{\frac{2t'^2}{\tau^2}}\left[\int_{-\infty}^{t'} e^{i\left[\left(\frac{-2\hbar k^2}{m}-2\omega\right)t''\right]} e^{\frac{t''^2}{\tau^2}} dt''\right] dt' + \int_{-\infty}^{\infty} e^{i\left[\left(\frac{2\hbar k^2}{m}-2\omega\right)t'\right]} e^{\frac{t'^2}{\tau^2}}\left[\int_{-\infty}^{t'} e^{i\left[\left(\frac{-2\hbar k^2}{m}+2\omega\right)t''\right]} e^{\frac{2t''^2}{\tau^2}} dt''\right] dt'\right\}$$

$$-2\frac{q^3 A_0^3 N^{\frac{3}{2}} p_x}{16m^2\hbar^2}\left\{\int_{-\infty}^{\infty} e^{i\left[\left(\frac{3\hbar k^2}{2m}-\omega\right)t'\right]} e^{\frac{2t'^2}{\tau^2}}\left[\int_{-\infty}^{t'} e^{i\left[\left(\frac{-3\hbar k^2}{2m}+\omega\right)t''\right]} e^{\frac{t''^2}{\tau^2}} dt''\right] dt' + \int_{-\infty}^{\infty} e^{i\left[\left(\frac{3\hbar k^2}{2m}+\omega\right)t'\right]} e^{\frac{t'^2}{\tau^2}}\left[\int_{-\infty}^{t'} e^{i\left[\left(\frac{-3\hbar k^2}{2m}-\omega\right)t''\right]} e^{\frac{2t''^2}{\tau^2}} dt''\right] dt'\right\} \quad (A5)$$

In order to evaluate these integrals the approximation $\int_{-\infty}^{t} e^{i\Omega t'} e^{-\frac{t'^2}{\tau^2}} dt' \approx \frac{-i}{\Omega} e^{i\Omega t} e^{-\frac{t^2}{\tau^2}}$ for $\Omega\tau \gg 1$ was used. By applying this directly to the integral above the probability amplitude is obtained.

$$C_{fi} \approx \frac{-q^3 A_0^3 N^{\frac{3}{2}} p_x}{16m^2\hbar^2}\left\{-i7\sqrt{\frac{\pi}{3}} \frac{\tau}{\omega} \frac{\hbar k}{mc}\right\} = i\sqrt{\frac{\pi}{3}} \frac{7q^3 p_x \tau}{16\hbar m^3 c^2 \omega^3}\left(\frac{2I}{c\varepsilon_0}\right)^{\frac{3}{2}}. \quad (A6)$$